\documentclass[pra,reprint,floatfix,superscriptaddress]{revtex4-1}
\usepackage[pdftex]{graphicx} 
\usepackage{dcolumn} 
\usepackage{bm} 
\usepackage{amssymb} 
\usepackage{placeins}
\usepackage{amsmath} 
\usepackage{footmisc}
\usepackage[latin1]{inputenc}
\usepackage{tikz}
\usetikzlibrary{shapes,arrows}
\usepackage[hidelinks]{hyperref}
\newcommand{\footnoteremember}[2]{
\footnote{#2}
\newcounter{#1}
\setcounter{#1}{\value{footnote}}}

\begin{document}

\title{Accurate real-time evolution of electron densities and ground-state properties from generalized Kohn-Sham theory}

\author{M.\ J.\ P.\ Hodgson}
\email[Personal email: ]{mhodgson@mpi-halle.mpg.de}
\homepage[Personal webpage: ]{http://www-users.york.ac.uk/~mjph501/}
\thanks{These two authors contributed equally}
\affiliation{Max-Planck-Institut f\"ur Mikrostrukturphysik, Weinberg 2, D-06120 Halle, Germany}
\affiliation{European Theoretical Spectroscopy Facility}
\author{J.\ Wetherell}
\email[Personal email: ]{jack.wetherell@polytechnique.edu}
\homepage[Personal webpage: ]{http://www-users.york.ac.uk/~jw1294/}
\thanks{These two authors contributed equally}
\affiliation{LSI, \'Ecole Polytechnique, CNRS, Institut Polytechnique de Paris, F-97728 Palaiseau}
\affiliation{European Theoretical Spectroscopy Facility}

\date{\today}
\begin{abstract}
The exact static and time-dependent Kohn-Sham (KS) exchange-correlation (xc) potential is extremely challenging to approximate as it is a local multiplicative potential that depends on the electron density everywhere in the system. The KS approach can be generalised by allowing part of the potential to be spatially nonlocal. We take this nonlocal part to be that of unrestricted Hartree-Fock theory. The additional local correlation potential in principle ensures that the single-particle density exactly equals the many-body density. In our case, the local correlation potential is predominantly nearsighted in its dependence on the density and hence an (adiabatic) local density approximation to this potential yields accurate ground-state properties and real-time densities for one-dimensional test systems.
\end{abstract}

\maketitle

\section{Introduction}

Models which reliably describe excited many-body systems for a low computational cost have remained elusive within solid state physics and quantum chemistry, despite their importance. Density functional theory\cite{PhysRev.136.B864} (DFT), within the Kohn-Sham (KS) approach\cite{Kohn-Sham}, is an extremely popular method for ground-state electronic structure calculations owing to its computationally efficiency and accuracy for most solids\cite{Bowler_2016}. However, for modeling systems with strong electron localization, such as molecules, the exact multiplicative exchange-correlation (xc) potential of KS theory has been shown to exhibit important features which have a strong nonlocal dependence on the density\cite{perdew1982density,almbladh1985density,PhysRevA.40.4190,:/content/aip/journal/jcp/131/22/10.1063/1.3271392,PhysRevA.83.062512,burke2012perspective,PhysRevB.93.155146}, which common approximations fail to capture\cite{zhang1998challenge,:/content/aip/journal/jcp/125/20/10.1063/1.2403848,:/content/aip/journal/jcp/125/19/10.1063/1.2387954,sousa2007general,cohen2008insights,:/content/aip/journal/jcp/129/12/10.1063/1.2987202,PhysRevLett.100.146401,mori2014derivative}, e.g., sudden changes in the level of the potential termed `steps' and `peaks'\cite{van1995step}. These missing features lead to an inaccurate description of the system, e.g., when the atoms of a diatomic molecule are dissociated\cite{cohen2008insights}. Features of the exact KS potential which are absent from approximations have a further adverse affect on the reliable prediction of excitation properties, such as excitation energies\cite{PhysRevB.95.035120,elliott2011perspectives,isegawa2012performance,hodgson2017interatomic} and real-time densities\cite{PhysRevLett.95.203004,PhysRevLett.104.236801,fuks2013dynamics,PhysRevB.93.155146}.

Accurate dynamic densities are crucial for predicting currents and the electronic properties of molecules which are perturbed by an external field. In particular, when acting as a molecular junction, the system is well beyond the linear-response and steady-state regimes. Hence commonly used approximations within DFT predict current-voltage characteristics within molecular electronic systems which are incorrect by orders of magnitude\cite{PhysRevB.69.235411}. Time-dependent DFT\cite{PhysRevLett.52.997} (TDDFT) is in principle a powerful method for describing excited systems. Standard approximations, e.g., the adiabatic local density approximation (ALDA) and adiabatic generalized gradient approximation\cite{PhysRevLett.77.3865}, have proved successful within KS TDDFT\cite{burke2005time}, e.g., within linear response for predicting photo-absorption spectra\cite{gross2012introduction}, although are less reliable in the presence of charge transfer\cite{dreuw2004failure,PhysRevB.86.201109,fuks2014charge,maitra2017charge} or when strong currents flow\cite{petersilka1999strong,Koentopp_2008,PhysRevLett.101.166401,PhysRevLett.104.236801,PhysRevB.88.241102,chirilua2017time,jornet2019real} owing to the absence of features which have a nonlocal dependence on the density\cite{PhysRevLett.89.023002,PhysRevLett.109.266404,PhysRevLett.108.146401,ramsden2012exact}. Novel functionals\cite{petersilka1999strong,PhysRevB.90.241107,chirilua2017time} or spin symmetry breaking\cite{PhysRevA.83.042501} have been employed in order to improve the reliability of these calculations. However, advanced approximations are required for real-time TDDFT to become as generally reliable as ground-state DFT\cite{provorse2016electron}.

The use of hybrid functionals\cite{becke1993new} within (TD)DFT has overcome some of the issues which face approximations within standard KS theory owing to the inclusion of a spatially nonlocal potential within the auxiliary KS system\cite{heyd2003hybrid,stein2009reliable,autschbach2009charge,kummel2017charge,PhysRevLett.119.145501,zhang2020linear}, e.g., the calculation of the fundamental gap\cite{muscat2001prediction,jain2011reliability}. Hybrid functionals offer a balance between accuracy and computational efficiency, e.g., for calculating quasiparticle energies and ground-state densities\cite{PhysRevMaterials.2.040801}. They have also been employed to tackle the challenging task of modeling dynamic systems\cite{lopata2011modeling}, however usually require empirical parameters. Hybrid functionals exist within the framework of generalized Kohn-Sham (GKS) theory\cite{seidl1996generalized}. GKS theory establishes that for a given spatially nonlocal potential there exists a unique spatially local (multiplicative) potential which ensures the single-particle density exactly equals the many-body density\cite{seidl1996generalized}. The form of this multiplicative potential thus depends on the choice of nonlocal potential. Our aim is to obtain a multiplicative potential which possess Kohn's concept of `nearsightedness'\cite{PhysRevLett.76.3168,Prodan11635}. The standard KS potential does not have this advantageous property as it depends on the density everywhere in the system\cite{PhysRevB.99.045129}. A nearsighted potential is concerned only with the properties of the system in its \textit{local} vicinity, and hence is in principle more accurately approximated on the basis of the local and semi-local density. This principle of nearsightedness may also be applied to time-dependent systems -- GKS theory has recently been extended to systems undergoing excitations\cite{kummel2017charge,baer2018time}.

In this paper we consider two choices for the nonlocal potential. The first is that of restricted Hartree-Fock (RHF) theory, which leads to a set of RHF-Kohn-Sham (RHFKS) equations within GKS theory, derived in Ref.~\onlinecite{seidl1996generalized}. The second is that of \textit{unrestricted} Hartree-Fock (UHF) theory\cite{PhysRev.102.1303}, which leads to a set of UHF-Kohn-Sham (UHFKS) equations, which we derive below. 

\section{Unrestricted Hartree-Fock-Kohn-Sham theory}

Within GKS theory the electrons must be described by a single Slater determinant (SD). We employ an `unrestricted SD', $\Phi$, in which electrons with different spins occupy different single-particle orbitals, i.e., $\psi_k(x,\sigma) = \phi_k(x) \gamma_k(\sigma)$ where $\gamma_{\cdot} = \alpha$ for up-spin electrons and $\gamma_{\cdot} = \beta$ for down-spin electrons\footnote{$\alpha(\sigma=\tfrac{1}{2}) = 1$, $\alpha(\sigma=-\tfrac{1}{2}) = 0$, $\beta(\sigma=\tfrac{1}{2}) = 0$ and $\beta(\sigma=-\tfrac{1}{2}) = 1$.}. This is in contrast to the SD of RHF in which two electrons with opposite spins occupy the same orbital. We then define the functional
\begin{equation} \label{eq:S}
    S[\Phi] = \left \langle \Phi \left | \hat{T} + \hat{U}(x-x') \right | \Phi \right \rangle,
\end{equation}
where $\hat{T}$ is the kinetic energy operator and $\hat{U}$ is the electron-electron interaction operator. From Eq.~(\ref{eq:S}) a unique density functional can be defined via the constrained search formalism of DFT\cite{levy1979universal}:
\begin{equation} \label{S}
Q^S[n] = \min_{\Phi \rightarrow n} S[\Phi],
\end{equation}
where the minimization searches over all SDs that yield the electron density $n$. This then allows us to define our correlation energy functional as the difference between $Q^S[n]$ and the Hohenberg-Kohn functional (which exactly captures all xc effects)\cite{PhysRev.136.B864}:
\begin{equation} \label{eq:Ec}
E_\mathrm{c}[n] \equiv \left \langle \Psi[n] \left | \hat{T} + \hat{U} \right | \Psi[n] \right \rangle - Q^S[n],
\end{equation}
where $\Psi[n]$ is the ground-state many-body wavefunction which yields the electron density $n$. Because UHF captures exchange and static correlation effects\cite{generalizedHF}, the correlation energy defined by Eq.~(\ref{eq:Ec}) approximately corresponds to `dynamic correlation'. 

The exact total ground-state many-body energy can be written in terms of these functionals, as such
\begin{widetext}
\begin{equation}
E_0 = \min_{\{ \phi_k \} \rightarrow N} \left \{ S[\{ \phi_k \}] + E_\mathrm{c}[n[\{ \phi_k \}]] + \int \mathrm{d} x \ v_{\mathrm{ext}}(x) n([\{ \phi_k \}];x) \right \},
\end{equation}
\end{widetext}
where $v_\mathrm{ext}$ is the external potential of the many-body system. Finally the set of single-particle UHFKS equations can be derived via a minimization of this energy employing Lagrangian multipliers to ensure orthogonality of the single-particle orbitals, $\{ \phi_k \}$:
\begin{widetext}
\begin{equation}
    \left( -\frac{1}{2} \frac{\mathrm{d}^2}{\mathrm{d}x^2} + v_\mathrm{ext}(x) + v_\mathrm{H}(x) + v_\mathrm{c}[n](x)\right)\phi^\gamma_i(x) + \int \mathrm{d}x' F_\mathrm{x}^\gamma(x,x') \phi^\gamma_i(x') = \varepsilon^\gamma_i \phi^\gamma_i(x),
    \label{eq:UHFKS}
\end{equation}
\end{widetext}
(atomic units are used throughout: $\hbar = e = m_\mathrm{e} = 4 \pi \varepsilon_0 = 1$) where the spatially nonlocal Fock opertor is 
\begin{equation} \label{eq:Fock}
    F_\mathrm{x}^\gamma(x,x') = -\sum^\mathrm{occ}_{j} \left( \phi_j^\gamma (x) \right)^* \phi^\gamma_j(x') U(x-x'),
\end{equation}
and
\begin{equation}
    v_\mathrm{c}[n](x) = \frac{\delta E_\mathrm{c}[n]}{\delta n(x)},
\end{equation}
which is a unique functional of the density. All electrons experience the same multiplicative Hartree potential, $v_\mathrm{H}$, and multiplicative local correlation potential, $v_\mathrm{c}$. In principle the single-particle density given by $\sum_{i, \gamma} | \phi^\gamma_i(x) |^2$ exactly equals the many-body density, $n(x)$; however, in practice the local correlation potential, to which we now turn our attention, must be approximated.

\section{Model systems}

We model ground-state and time-dependent one-dimensional (1D) systems consisting of two opposite-spin electrons. As a result, our examples do not illustrate the exchange interaction between pairs of like-spin electrons, which would be captured by our exchange potential, but instead focus on the more challenging aspect of accurately approximating correlation. Nonetheless, our use of a nonlocal exchange potential is necessary as it induces nearsightedness in the corresponding multiplicative potential when the system is comprised of like-spin electrons, demonstrated in Ref.~\onlinecite{PhysRevB.99.045129}, e.g., when the system consists of more than two electrons.

We focus on these small yet challenging model systems as all the exact quantities can be numerically determined, including the fully-correlated many-body wavefunction, by solving the ground-state or time-dependent many-body Schr\"odinger equation for any external potential, $v_\mathrm{ext}(x,t)$, with the appropriately softened 1D Coulomb interaction $(\left|x-x' \right|+1)^{-1}$\cite{PhysRevA.72.063411} by employing our \texttt{iDEA} code\cite{PhysRevB.88.241102}. 

First we use our exact solutions to the Schr\"odinger equation to find the \textit{exact} local correlation potential of RHFKS and UHFKS theory. We then develop an LDA to each of these correlation potentials, termed RLDA+ and ULDA+ to distinguish them from the usual LDA employed within standard KS theory. We assess their accuracy for various ground-state and time-dependent systems. We also model our systems with (time-dependent) RHF and UHF theory as a benchmark.

\subsection{The exact local correlation potentials} \label{sec:exact_vc}

We calculate the exact $v_\mathrm{c}(x)$ of RHFKS theory and that of UHFKS theory for ground-state 1D models of the hydrogen molecule (H$_2$). As we have access to the exact many-body density we can `reverse-engineer' the RHFKS and UHFKS equations in order to find the corresponding exact correlation potentials for each system.

\begin{figure}[htbp] 
    \centering
    \includegraphics[width=1.0\linewidth]{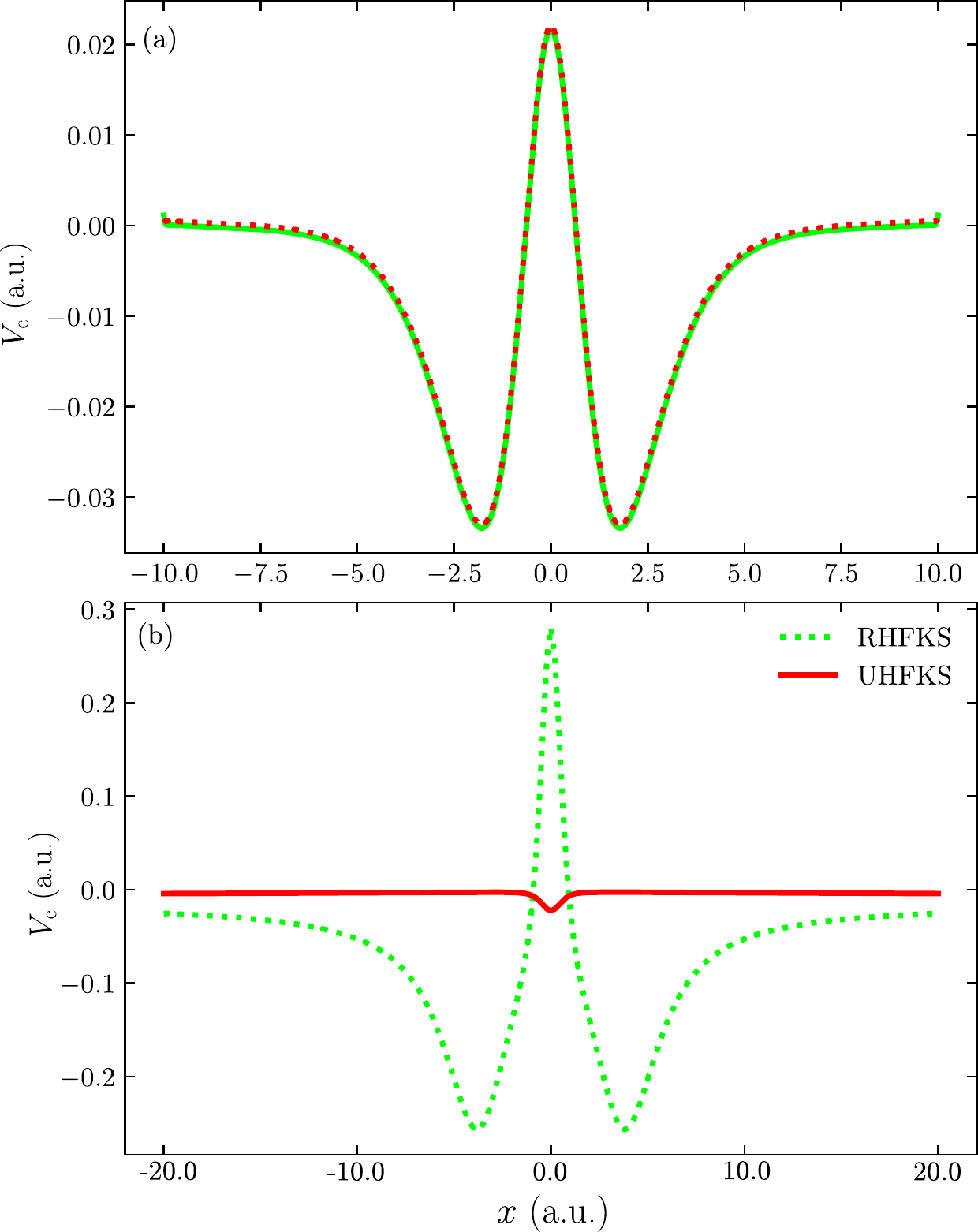}
    \caption{(a) The correlation potential of RHFKS and UHFKS at the bonding length (atomic separation $\approx 1.4$ a.u.) -- they are indistinguishable. (b) The same potentials when the atoms are dissociated (atomic separation = 7.2 a.u.). The correlation potential of RHFKS has a nonlocal dependence on the density -- there is a peak between the electrons in a region of very low density; see Fig.~\ref{fig:H2_den_EXT_HF_UHF}(b). The correlation potential of UHFKS tends to a constant as the atoms are separated and has a more local dependence on the density.}
    \label{fig:H2_diss_corr_pot}
\end{figure}

Figure~\ref{fig:H2_diss_corr_pot}(a) shows the exact correlation potential of RHFKS and UHFKS for H$_2$ at the bonding length (atomic separation $\approx 1.4$ a.u. -- calculated via the bonding energy; see Fig.~\ref{fig:H2_diss_E_EXT_HF_UHF}). The two potentials are indistinguishable, indicating the lack of static correlation at this atomic separation. As the atoms are dissociated these two correlation potentials diverge starting at the point where UHF breaks the spin symmetry (Coulson-Fischer point)\footnoteremember{ftn:SM_movie}{An animation which shows the adiabatic separation of the atoms and the affect on the corresponding correlation potential can be seen in our Supplemental Material}. 

\begin{figure}[htbp]
    \centering
    \includegraphics[width=1.0\linewidth]{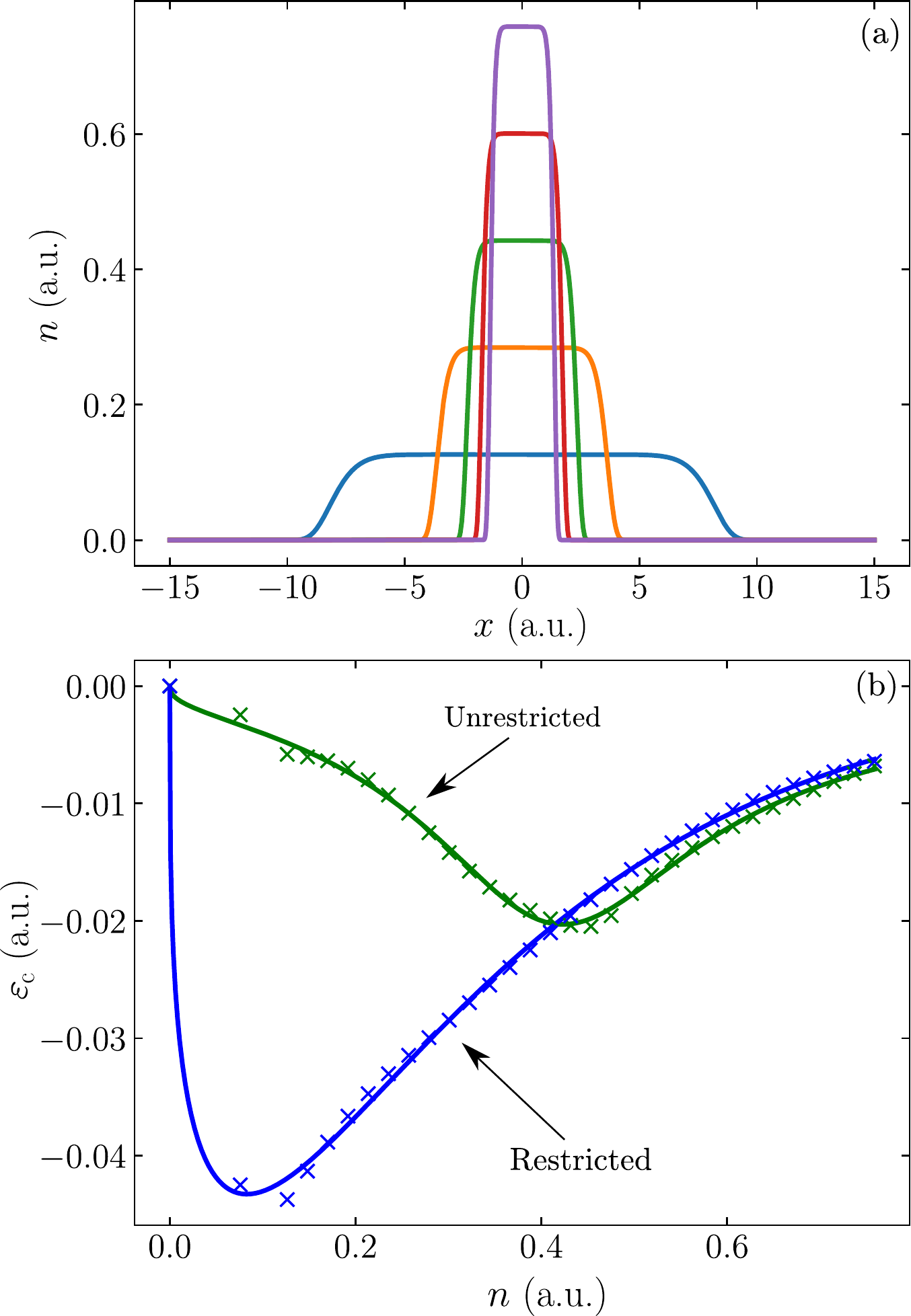}
    \caption{(a) The range of `slab densities' used to construct our RLDA+ and ULDA+. (b) The correlation energy per electron, which corresponds to UHFKS and RHFKS, of each slab against the density height of the corresponding slab density.}
    \label{fig:lda}
\end{figure}

The form of the exact $v_\mathrm{c}(x)$ of RHFKS has a nonlocal dependence on the density. This multiplicative potential distributes the electron density throughout the system in a similar fashion to the xc potential of standard KS theory (for an asymmetric system the peak in the center would be a step\cite{PhysRevB.93.155146}; see Fig.~\ref{fig:H2_diss_corr_pot}(b) and the Supplemental Material). These features are notoriously difficult to capture in approximate methods. Dreissigacker and Lein demonstrated that by breaking spin symmetry, as we have done here, one may capture some of these features in a multiplicative potential, but such an approach requires a spin-DFT calculation and then an expensive numerical inversion of the yielded density\cite{dreissigacker2011adiabatic}. Our goal is to find a local correlation potential which is accurately represented by a local or semi-local approximation, and hence our ideal multiplicative potential should be free from features with a strong nonlocal dependence on the density.

Figure~\ref{fig:H2_diss_corr_pot} shows that the exact correlation potential of UHFKS does \textit{not} contain prominent features with a strong nonlocal dependence on the density. This stems from the use of different orbitals for different spins: Because each electron is free to independently experience the Hartree potential of the other electron without experiencing its own Hartree potential, there is no need for the peak (or step) in the multiplicative potential in order to correctly distribute the density throughout the system when the atoms are dissociated. In effect each electron is able to experience its own effective potential which renders the need for a peak or step in the local correlation potential unnecessary. This is demonstrated for like-spin electrons in Ref.~\onlinecite{PhysRevB.99.045129}. 

\subsection{Constructing an LDA for each local correlation potential}

Next we construct an LDA to the correlation potential of RHFKS and UHFKS, termed RLDA+ and ULDA+ respectively. We use a set of `slab systems'\cite{PhysRevB.94.205134} which are homogeneous in the center of the system and tend to zero towards the edge, as in Ref.~\onlinecite{PhysRevB.94.205134}. We vary the height of the slabs ($n$) and calculate the correlation energy ($E_\mathrm{exact}-E_\mathrm{HF}$) for each system\footnote{This way of constructing an LDA was done for the xc of KS theory and shown to yield an LDA close to that compared to the LDA constructing from the homogeneous electron gas in Ref.~\onlinecite{PhysRevB.97.235143}.}; see Fig.~\ref{fig:lda}. Once an LDA for each correlation energy is constructed we apply it to its own training systems and find small errors in $E_\mathrm{c}$ ($\Delta E_\mathrm{c}$) owing to the inhomogeneity of the slab systems. We use these errors to calculate a refined LDA+: $\varepsilon_\mathrm{c}(n) \rightarrow \varepsilon_\mathrm{c}(n) - \Delta E_\mathrm{c}(n)/2$.

Figure~\ref{fig:lda}(b) shows the correlation energy per electron for the range of slab systems ($\varepsilon_\mathrm{c}$) as a function of the density, $n$. The form of the `restricted correlation energy', i.e., that which corresponds to RHFKS, is similar to the `unrestricted correlation energy' for high density regions because for these systems static correlation is negligible. However, for the lower density regions the two energies differ to a large degree. Note that the restricted correlation energy is much larger than the unrestricted correlation energy because it has the burden of capturing static correlation\footnote{The analytic formulas for each correlation energy can be found in our Supplemental Material.}.

\subsection{Ground-state hydrogen molecule}

\begin{figure}[htbp]
    \centering
    \includegraphics[width=1.0\linewidth]{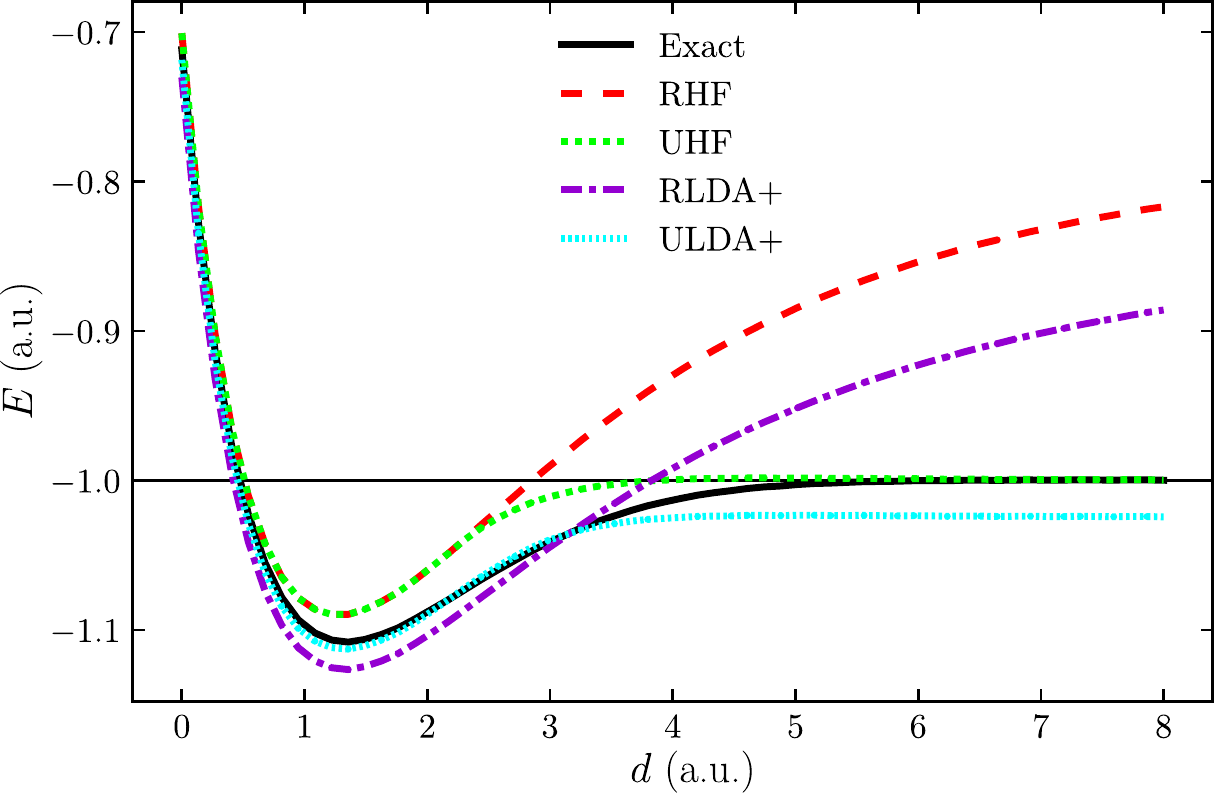}
    \caption{Molecular energy of H$_2$ as the atoms are separated. RHF and UHF are compared against the many-body exact. As expected, RHF is incorrect for all separations and UHF is inaccurate at and around the bonding length owing to an absence of dynamical correlation effects. Our RLDA+ only slightly improves the energy. Our ULDA+ gives a large improvement upon UHF at and around the bonding length and slightly worsens the performance of UHF in the dissociation limit.}
    \label{fig:H2_diss_E_EXT_HF_UHF}
\end{figure}

We now calculate the molecular energy of H$_2$ as the atoms are separated employing RHF, UHF, RLDA+ and ULDA+ and compare them to the exact case; for a comparison of our 1D results against the corresponding experimental results see our Supplemental Material. Figure~\ref{fig:H2_diss_E_EXT_HF_UHF} shows, as expected, that the RHF energy is inaccurate owing to a complete absence of correlation. UHF correctly gives the dissociation energy of the molecule by capturing static correlation\cite{generalizedHF} but yields the incorrect energy at and around the bonding length ($\approx 1.4$ a.u.) owing to the absence of dynamic correlation in the approximation. At the bonding length RHF and UHF yield exactly the same total electron energy with an error of $1.3\%$. The error in the ionization potential (IP) predicted by UHF and RHF is $2.1\%$. At an atomic separation of $7.2$ a.u.\ the error in the IP of RHF is $34.3\%$ and $1.7\%$ for UHF. 

ULDA+ yields an accurate total electron energy at the bonding length with an error of $0.5 \%$ and introduces an error as the atoms are separated ($2 \%$) because of the use of an LDA to $v_\mathrm{c}$; as the electrons localize each to an H atom, our approximate correlation potential introduces a `self-correlation error'. In principle this error could be reduced via a more sophisticated approximation to $v_\mathrm{c}$\cite{PhysRevB.97.121102}.

RLDA+ gives a relatively poor total energy at the bonding length, $1.6\%$, and an only slightly improved total energy when the molecule is stretched, $9.0\%$, when compared to the RHF error of $15.2\%$. Because the LDA for the restricted correlation energy aims to add both static and dynamic correlation, and there is negligible static correlation in the systems with relatively small atom separations, RLDA+ yields relatively inaccurate energies by introducing spurious static correlation. This issue is not present for ULDA+ as it only attempts to introduce dynamic correlation which is present in these systems\footnote{Note that within GKS theory the total electron energy is calculated via the appropriate energy density functional, as is appropriate within DFT.}.

\begin{figure}[htbp]
    \centering
    \includegraphics[width=1.0\linewidth]{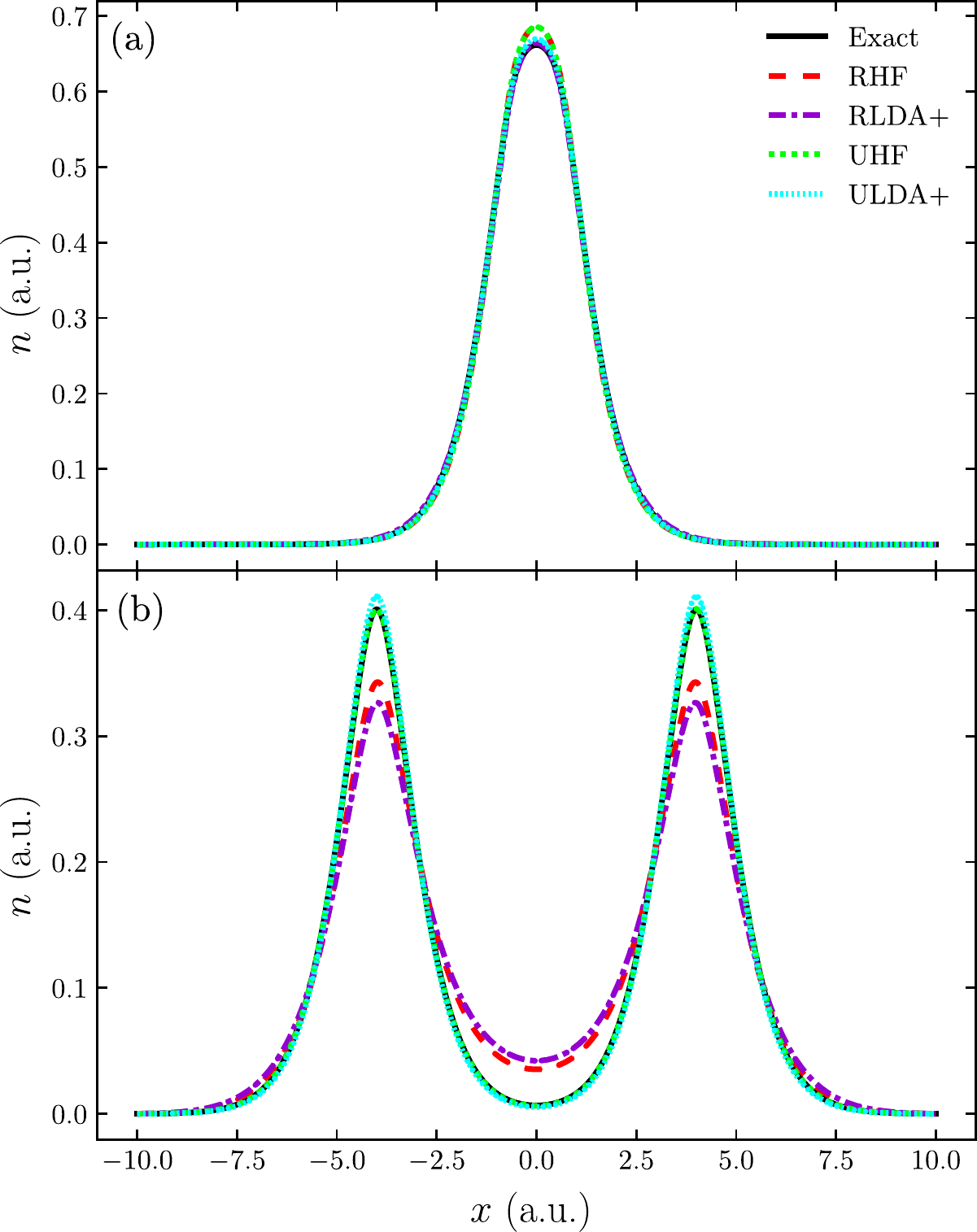}
    \caption{(a) The H$_2$ molecule at the bonding length (atomic separation $\approx 1.4$ a.u.). UHF and RHF densities are indistinguishable: both are inaccurate in the central region when compared to the many-body exact. The decay of the approximate densities is also incorrect. RLDA+ and ULDA+ yield accurate densities. (b) The dissociated H$_2$ molecule (separation $=8.0$ a.u.). UHF density is extremely accurate. The RHF density is poor when compared to the exact owing to the absence of static correlation. RLDA+ does not improve the density much over RHF, showing the in inability to capture static correlation with an LDA. The ULDA+ density is only slightly worse than UHF owing to the self-correlation error (see text) introduced by the local approximation to the correlation energy.}
    \label{fig:H2_den_EXT_HF_UHF}
\end{figure}

Next we turn to the ground-state density. We compare the density from RLDA+ and ULDA+ to RHF, UHF and the exact at the bonding length (see Fig.~\ref{fig:H2_den_EXT_HF_UHF}(a)) and when the atoms are dissociated; see Fig.~\ref{fig:H2_den_EXT_HF_UHF}(b). When the atoms are relatively close together, RHF and UHF yield the same density (like for the energy). RLDA+ and ULDA+ also yield very similar densities, with ULDA+ performing slightly better. 

On the other hand, when the atoms are separated the RHF and RLDA+ densities are inaccurate. RLDA+ only slightly improves the electron density compared to RHF demonstrating how ineffective an LDA to the restricted correlation potential is, as expected; see Sec.~\ref{sec:exact_vc}. UHF is extremely accurate when the atoms are dissociated, again as expected. ULDA+ is only slightly worse for the density demonstrating that the self-correlation error introduced by our LDA to the correlation potential does not have a detrimental effect.

\subsection{Time-dependent perturbed hydrogen molecule}

We now assess the accuracy of employing RLDA+ and ULDA+ adiabatically to calculate the time-dependent density and current when the H$_2$ molecule is perturbed by an electric field ($-0.03x$). The time-dependent version of Eq.~(\ref{eq:UHFKS}) is
\begin{equation}
    \hat{h}^\gamma \phi^\gamma_i(x,t) = i \frac{\partial}{\partial t} \phi^\gamma_i(x,t),
    \label{eq:TDUHFKS}
\end{equation}
where $\hat{h}^\gamma$ is the single-particle Hamiltonian given by Eq.~(\ref{eq:UHFKS}) but with the time-dependent potentials. (We do not derive this equation as the existence of a unique multiplicative potential cannot be ensured in general\cite{baer2018time}, but is instead an ansatz.)

\begin{figure}[htbp]
    \centering
    \includegraphics[width=1.0\linewidth]{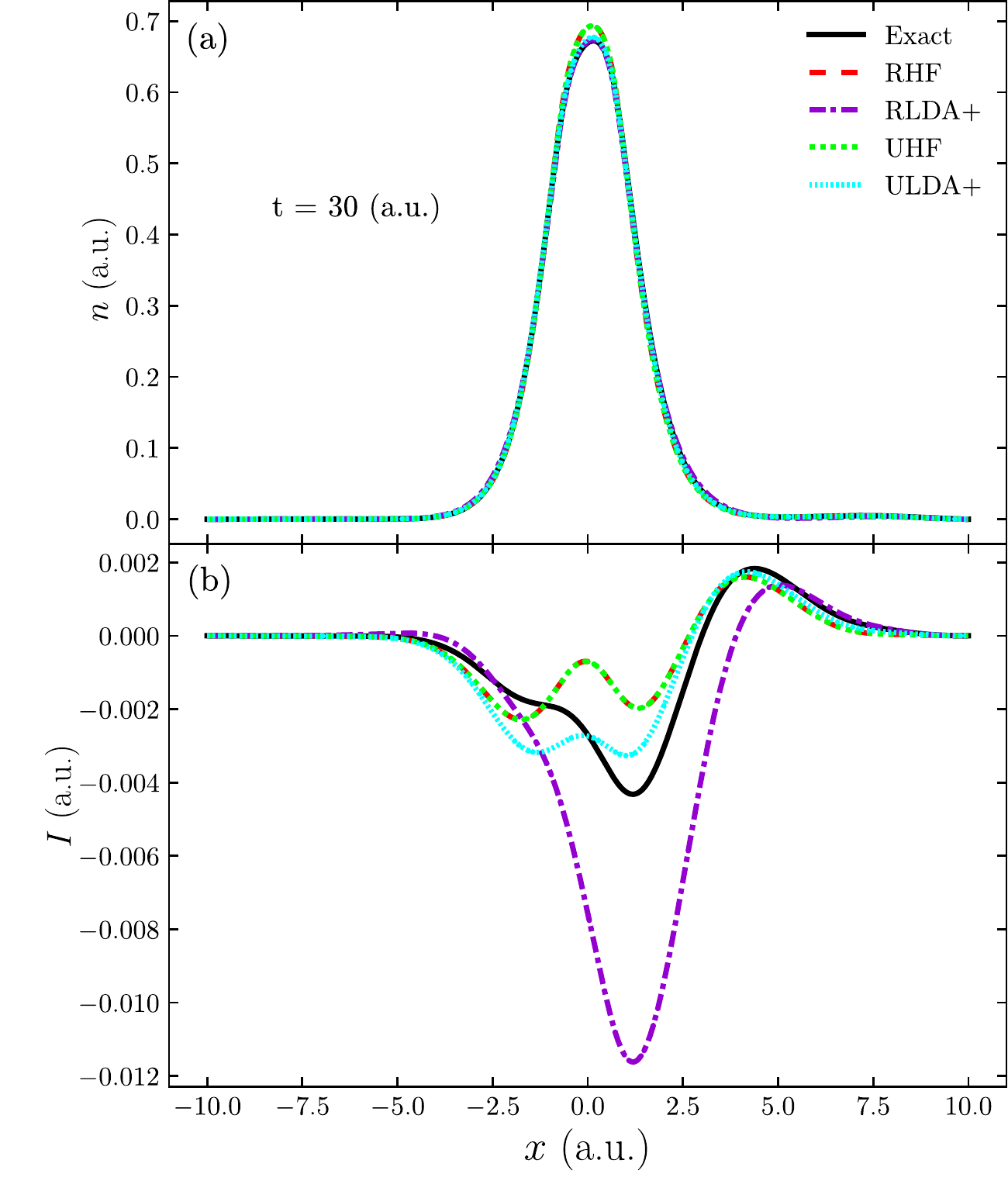}
    \caption{(a) The dynamic electron density at $t=30$ a.u. All approximations yield similar densities by eye. (b) The corresponding electron current. The performance of each approximation is more apparent. RLDA+ yield by far the worst result, while ULDA+ the best. Both RHF and UHF yield relatively accurate currents.}
    \label{fig:H2_TD_den_EXT_HF_UHF}
\end{figure}

\begin{figure}[htbp]
    \centering
    \includegraphics[width=1.0\linewidth]{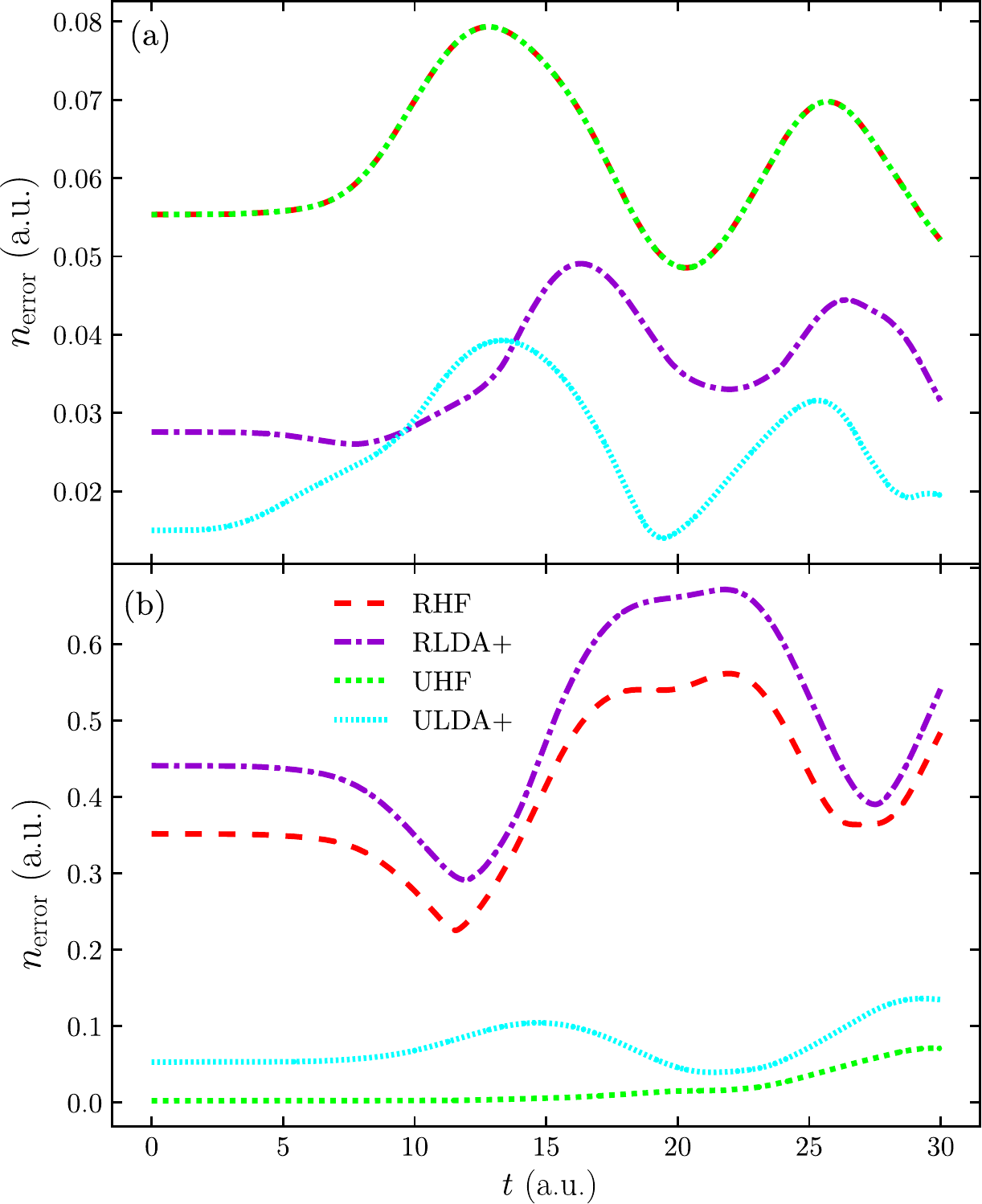}
    \caption{The integrated absolute density error of each approximation as a function of time. (a) Applied to the molecule at the bonding length ($\approx 1.4$ a.u.). The RHF and UHF methods are indistinguishable in this case, as expected. The ULDA+ method successfully reduces the error relative to UHF and approximately reduces the error by a further factor of two when compared to RLDA+. (b) Applied to the molecule at the stretched length (8.0 a.u.). The UHF method outperforms the RHF method. The RLDA+ and ULDA+ methods perform slightly worse than their counterparts due to the presence of small spurious self-interaction arising from the LDA.}
    \label{fig:TD_ERR}
\end{figure}

Figure~\ref{fig:H2_TD_den_EXT_HF_UHF}(a) shows the electron density at $t = 30$ a.u.\ for the hydrogen molecule at the bonding length after the perturbation. The density sloshes back and forth within the molecule and generates a current; see Fig.~\ref{fig:H2_TD_den_EXT_HF_UHF}(b). The current is the clearest indicator of how accurate each approximation is: RLDA+ is  the worst, RHF and UHF perform identically and yield a reasonably good approximation to the many-body electron current. The best method is ULDA+: although the current is by no means exact, it is quantitatively correct throughout the whole simulation ($30$ a.u.). The relative error of the time-dependent densities as the system evolves is shown in Fig.~\ref{fig:TD_ERR}(a).

\begin{figure}[htbp]
    \centering
    \includegraphics[width=1.0\linewidth]{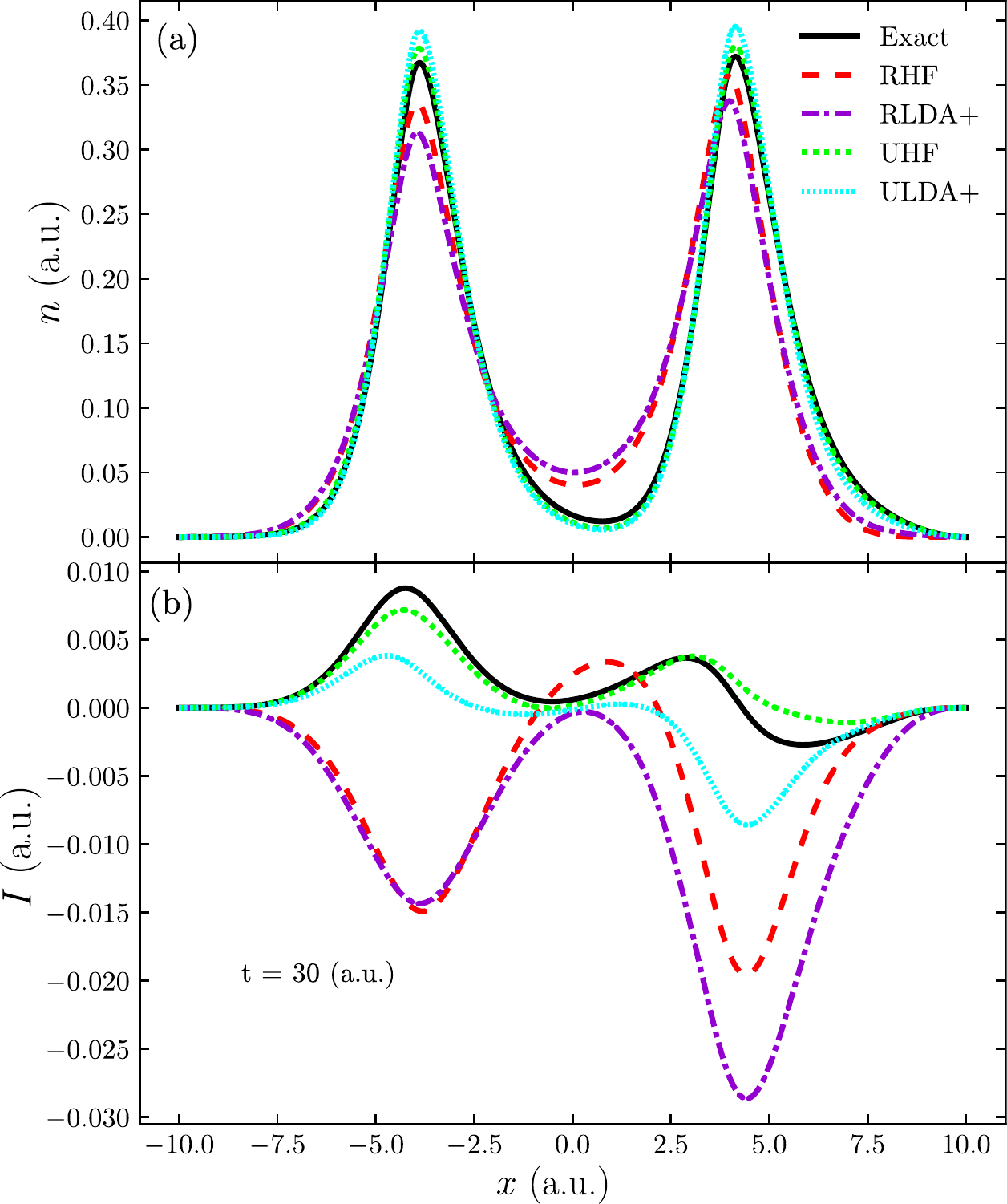}
    \caption{(a) Dynamic electron densities at $t=30$ a.u. The RHF and RLDA+ densities are visibly bad; whereas the UHF and ULDA+ densities are accurate throughout the whole simulation. (b) The same is the case for the current, with UHF yielding the most accurate current. ULDA+ yields a quantitatively good description of the current throughout the simulation.}
    \label{fig:DISS_H2_TD_den_EXT_HF_UHF}
\end{figure}

When the stretched molecule is perturbed, the story is similar: RLDA+ performs the worst, followed by RHF. However, for this system UHF is the most accurate followed by ULDA+. Like for the ground state, the approximation to the correlation potential has not significantly worsened the density when the ULDA+ is employed. Again, in principle this error can be reduced by more advanced (semi-local) approximations to the time-dependent correlation potential. Again see Fig.~\ref{fig:TD_ERR}(b) for the relative error of the time-dependent densities. (Videos of all time-dependent results can be found in the Supplemental Material.)

\section{Conclusion}

We have shown that, within generalized Kohn-Sham (KS) theory, the multiplicative local potential which ensures an exact density can be well approximated on the basis of the local density when the potential is designed to be `nearsighted' through the choice of spatially nonlocal potential. We find that the exact local correlation potential which corresponds to the nonlocal potential of unrestricted Hartree-Fock (UHF) theory does not contain features which depend on the density everywhere in the system, which are present in the exact exchange-correlation potential of standard KS theory, e.g., steps and peaks. On the other hand, the local correlation potential which corresponds to the nonlocal potential of restricted HF theory does have such features because, in this case, without spin symmetry breaking, the contribution from the nonlocal potential does not render the corresponding local potential nearsighted. 

We constructed a local density approximation (LDA) to the local correlation potential of UHFKS theory, which we term `ULDA+'. Our approximation yields accurate ground-state densities and energies for 1D model hydrogen molecules for varying atomic separations, and can even yield relatively accurate currents when applied adiabatically to systems which are perturbed by an electric field. We compared our calculations to the exact many-body solutions. 

Our ULDA+ has an inherent `self-correlation error'. We find that this leads to inaccuracies in the energies and densities of H$_2$ when its atoms are dissociated, although these errors are significantly small relative to other approximate methods. This error can in principle be reduced by a more advanced approximation to the correlation energy. Overall, our ULDA+ yields accurate dynamic densities and currents as well as ground-state properties even when correlation is strong. Future work should focus on the development of a 3D ULDA+ which may lead to accurate simulations of realistic molecules undergoing field-induced excitations.

\acknowledgements{We thank Lucia Reining for feedback and the University of York for computational resources.}

\bibliography{Bibtex}

\end{document}